\documentstyle[amssymb,12pt,epsf]{article}

\newlength{\absize}
\input{tcilatex}

\begin{document}

\renewcommand{\title}[1]{    \begin{center}
\LARGE #1
\end{center}\par} 
\renewcommand{\author}[1]{    \vspace{2ex}
{\normalsize
 \begin{center}
\setlength{\baselineskip}{3ex} #1 \par
\end{center}}} \renewcommand{\thanks}[1]{\footnote{#1}} 
\renewcommand{\abstract}[1]{    \vspace{2ex}
\normalsize
\begin{center}
\centerline{\bf Abstract}\par
\vspace{2ex}
\parbox{\absize}{#1\setlength{\baselineskip}{2.5ex}\par}
\end{center}}

\begin{titlepage}
 $~~~~~~~~~~~~~~~~~~~~~~~~~~~~~~~~~~~~~~~~~~~~~~~~~~~~~~~$  CERN-TH/2000-309 
\vspace{3mm}

\title{A sum rule from the shape function}

\vskip 0.6cm

\begin{center}
{\large Ugo Aglietti$\; {}^{a}$\footnote{On leave of absence from Dipartimento di
Fisica, Universit\`a  di Roma I, Piazzale Aldo Moro 2, 00185, Roma; \sl e-mail: 
ugo.aglietti@cern.ch} }

\vskip 0.5cm

$^a${\sl
TH Division, CERN, Geneva, Switzerland }



\end{center}

\abstract{ We present a sum rule relating
the electron energy spectrum to the hadron
mass distribution in semileptonic $b\rightarrow u$ decays close to
threshold. The relation found is free from non-perturbative effects in leading twist, so the
theoretical error is expected to be $O\left( 5\%\right) $. An experimental confirmation
of this prediction can provide a check of the basic assumptions at the root of
the theory of the shape function.  }

\vskip 7 cm

\noindent

CERN-TH/2000-309

October 2000
\end{titlepage}

\setcounter{footnote}{0}

\newpage

In this note we present a sum rule that can be directly compared with data
on the semileptonic decay 
\begin{equation}
B\rightarrow X_{u}+l+\nu .  \label{inizio}
\end{equation}
The comparison allows a verification of the theory of the structure function
for the heavy flavours, usually called the shape function \cite{generale,
noi2}. This sum rule relates the electron spectrum to the integrated (also
called cumulative) hadron-mass distribution and reads 
\begin{equation}
\frac{d\Gamma _{B}}{dx}=2C\left( \alpha _{S}\right) \int_{0}^{m_{B}\sqrt{1-x}%
}\frac{d\Gamma _{B}}{dm_{X}}\,dm_{X}\,\,+\,O\left( \frac{\Lambda _{QCD}}{%
m_{B}}\right) ,  \label{lei}
\end{equation}
where the coefficient function is, to one loop, 
\begin{equation}
C\left( \alpha _{S}\right) =1+\frac{C_{F}\,\alpha _{S}}{2\pi }\,\frac{97}{72}%
+O\left( \alpha _{S}^{2}\right) .
\end{equation}
The adimensional electron energy is defined, as usual, as 
\begin{equation}
x\equiv \frac{2E_{e}}{m_{B}}\qquad \qquad \left( 0\leq x\leq 1\right) .
\end{equation}
Relation (\ref{lei}) holds in the region 
\begin{equation}
1-x\,\sim \,\frac{\Lambda _{QCD}}{m_{B}}.  \label{cond}
\end{equation}
Assuming $\Lambda _{QCD}\sim 300$ MeV, this means \footnote{%
In practice, to kill the large $b\rightarrow cl\nu $ background, one has to
satisfy the experimental constraint \cite{ACCMM} \label{noterella} 
\begin{equation}
x>\frac{m_{B}^{2}-m_{D}^{2}}{2m_{B}}\simeq 0.88.  \label{vecchiaamica}
\end{equation}
} 
\begin{equation}
E_{e}\sim 2.5\,{\rm GeV.}
\end{equation}
The condition (\ref{cond}) corresponds to a final invariant hadronic mass in
the region \cite{generale, noi2} 
\begin{equation}
m_{X}\sim \sqrt{\Lambda _{QCD}\,m_{B}}\sim m_{X}\sim 1.3\,{\rm GeV.}
\label{thr}
\end{equation}
As  stated in eq.\thinspace (\ref{lei}), the sum rule holds only if the
upper invariant mass $m_{cut}$ in the hadron distribution is related to the
electron energy by 
\begin{equation}
m_{cut}=m_{B}\sqrt{1-x}.  \label{vincolo}
\end{equation}
A typical value for the experimental analysis is $m_{cut}=1.6$ GeV, for
which $E_{e}=2.4$ GeV. One can actually decrease the cut mass to something
like $m_{cut}=1.3$ GeV, for which $E_{e}=2.48$ GeV (the end-point is at $%
E_{e}^{\max }=2.64$\thinspace GeV).

The coefficient function has the numerical value 
\begin{equation}
C\left( \alpha _{S}\right) \cong 1.06
\end{equation}
for $\alpha _{S}\equiv \alpha _{S}\left( m_{B}\right) =0.2.$ Taking instead,
for example, $\alpha _{S}\equiv \alpha _{S}\left( \mu =m_{B}/2\right) =0.28,$
the coefficient function rises to $1.08,$ a $2\%$ variation: this can be
taken as a crude estimate of the higher-order terms, $\sim \left( \alpha
_{S}/\pi \right) ^{2}.$ In general, the main corrections to eq.\thinspace (%
\ref{lei}) originate from the so-called higher-twist effects, related to the
matrix elements of power-suppressed operators. Their size is \cite{generale,
noi2}, as anticipated: 
\begin{equation}
{\rm (higher\,\,twist\,\,contrib.)\,\sim \,}\frac{\Lambda _{QCD}}{m_{B}}%
\,\sim \,5\,\%.
\end{equation}

The proof of eq.\thinspace (\ref{lei}) is the following. Any distribution in
the threshold region (\ref{thr}) satisfies the factorization formula (for a
derivation, see for example \cite{noi2}) 
\begin{equation}
d\Gamma _{B}=\int_{0}^{m_{B}}dm_{\ast }\,\varphi \left( m_{\ast }\right)
\,d\Gamma _{\ast }+O\left( \frac{\Lambda _{QCD}}{m_{B}}\right) ,
\label{fact}
\end{equation}
where $d\Gamma _{\ast }$ is the distribution for a hypothetical heavy quark
with mass $m_{\ast }$ and $\varphi \left( m_{\ast }\right) $ is the shape
function in the notation of ref.\thinspace \cite{noi2}.

The tree-level electron spectrum is, close to the end-point: 
\begin{eqnarray}
\frac{1}{\Gamma _{0}}\frac{d\Gamma _{\ast }}{dx_{\ast }} &=&2x_{\ast
}^{2}\left( 3-2x_{\ast }\right) \,\theta \left( 1-x_{\ast }\right)  
\nonumber \\
&\cong &2\left[ 1-3\left( 1-x_{\ast }\right) ^{2}\right] \,\theta \left(
1-x_{\ast }\right) ,  \label{tree}
\end{eqnarray}
where 
\begin{equation}
x_{\ast }\equiv \frac{2E_{e}}{m_{\ast }}\qquad \qquad \left( 0\leq x_{\ast
}\leq 1\right) 
\end{equation}
and\footnote{%
The actual value of the heavy mass entering $\Gamma _{0}$ is irrelevant, as
this dependence cancels in taking the ratio of the widths (see later).} 
\begin{equation}
\Gamma _{0}\equiv \frac{G_{F}^{2}m_{b}^{5}|V_{ub}|^{2}}{192\pi ^{3}}\,.
\end{equation}
The term quadratic in $1-x_{\ast }$ in the last member of eq.\thinspace (\ref
{tree}) can be neglected because 
\begin{equation}
\left( 1-x_{\ast }\right) ^{2}\sim \left( \frac{\Lambda _{QCD}}{m_{B}}%
\right) ^{2}.
\end{equation}
Performing then the linearization and inserting the r.h.s. of eq.\thinspace (%
\ref{tree}) into eq.\thinspace (\ref{fact}), one obtains 
\begin{eqnarray}
\frac{1}{\Gamma _{0}}\frac{d\Gamma _{B}}{dx} &=&2\int_{2E_{e}}^{m_{B}}dm_{%
\ast }\varphi \left( m_{\ast }\right) \frac{m_{B}}{m_{\ast }}  \nonumber \\
&=&2\int_{2E_{e}}^{m_{B}}dm_{\ast }\varphi \left( m_{\ast }\right) +O\left( 
\frac{\Lambda _{QCD}}{m_{B}}\right) ,
\end{eqnarray}
where  eq.\thinspace (\ref{cond}) has been used in the last line. An
analogous factorization of the hadron-mass distribution gives 
\begin{equation}
\,\frac{d\Gamma _{B}}{dm_{X}^{2}}=\int_{0}^{m_{B}}dm_{\ast }\varphi \left(
m_{\ast }\right) \frac{d\Gamma _{\ast }}{dm_{X}^{2}}.
\end{equation}
At tree level, the parton distribution reads 
\begin{equation}
\frac{1}{\Gamma _{0}}\frac{d\Gamma _{\ast }}{dm_{X}^{2}}=\delta \left(
m_{X}^{2}+2E_{X}\left( m_{\ast }-m_{B}\right) \right) ,
\end{equation}
where $E_{X}$ is the final hadronic energy. The latter has a range, for
fixed $m_{X}^{2},$%
\begin{equation}
m_{X}\leq E_{X}\leq \frac{m_{B}}{2}\left( 1+\frac{m_{X}^{2}}{m_{B}^{2}}%
\right) .
\end{equation}
We now introduce an approximation analogous to the one leading to the
factorization in terms of the variable mass $m_{\ast };$ we set\footnote{%
This step is not very rigorous. The main justification for neglecting the
region $E_{X}\gtrsim m_{X}$ is that infrared logarithms turn out to cancel
in the coefficient function $C\left( \alpha _{S}\right) $ (see later). For a
general discussion on this point, see ref. \cite{montp2}.}: 
\begin{equation}
E_{X}\sim \frac{m_{B}}{2}.
\end{equation}
Integrating over $m_{X}^{2},$ we obtain for the cumulative hadron-mass
distribution 
\begin{equation}
\frac{1}{\Gamma _{0}}\int_{0}^{m_{cut}}\,\frac{d\Gamma _{B}}{dm_{X}}%
\,dm_{X}=\int_{m_{B}\left( 1-m_{cut}^{2}/m_{B}^{2}\right) }^{m_{B}}dm_{\ast
}\,\varphi \left( m_{\ast }\right) .
\end{equation}
Comparing the expressions for the two distributions and assuming
eq.\thinspace (\ref{vincolo}), we obtain the tree-level approximation to
eq.\thinspace (\ref{lei}), i.e. the equation with $\alpha _{S}=0$ in $%
C\left( \alpha _{S}\right) $.

Let us now discuss the one-loop correction to $C\left( \alpha _{S}\right) $.
As is well known, the coefficient function is independent of the external
state chosen to perform the computation. Therefore, instead of \ taking an
external state containing a $B$-meson, we can take an external state
containing an on-shell $b$-quark. Then, we just need to consider the decay
distributions at the parton level to order $\alpha _{S}.$ The electron
spectrum reads \cite{jezkuhn} 
\begin{eqnarray}
\frac{1}{2\,\Gamma _{0}}\frac{d\Gamma _{b}}{dx} &=&1-\frac{\alpha _{S}C_{F}}{%
2\pi }\left[ \log ^{2}\left( 1-x\right) +\frac{31}{6}\log \left( 1-x\right)
+\pi ^{2}+\frac{5}{4}+\right.   \nonumber \\
&&\qquad \qquad \qquad \qquad \qquad \qquad \qquad \left. \QATOPD. .
{\,}{\,}\quad \quad +O\left( 1-x\right) \right] .  \label{elespec1}
\end{eqnarray}
\ The mass distribution is \cite{ndf}: 
\begin{eqnarray}
\frac{1}{\Gamma _{0}}\frac{d\Gamma _{b}}{dm_{X}^{2}} &=&\,\,\delta \left(
m_{X}^{2}\right) \left[ 1-\frac{\alpha _{S}C_{F}}{2\pi }\left( \pi ^{2}+%
\frac{187}{72}\right) \right] + \\
&&-\frac{\alpha _{S}C_{F}}{2\pi }\left[ 2\left( \frac{\log \left[
m_{X}^{2}/m_{b}^{2}\right] }{m_{X}^{2}/m_{b}^{2}}\right) _{+}+\frac{31}{6}%
\left( \frac{1}{m_{X}^{2}/m_{b}^{2}}\right) _{+}+O\left( 1\right) \right] , 
\nonumber
\end{eqnarray}
where the plus prescription regularizes the infrared singularity at $%
m_{X}^{2}=0,$ which comes from real gluon emission. Integrating over the
mass, we obtain 
\begin{eqnarray}
\frac{1}{\Gamma _{0}}\int_{0}^{m_{cut}}\frac{d\Gamma _{b}}{dm_{X}}\,dm_{X}
&=&1-\frac{\alpha _{S}C_{F}}{2\pi }\left[ \log ^{2}\left( 1-z\right) +\frac{%
31}{6}\log \left( 1-z\right) +\right.   \nonumber \\
&&\qquad \qquad \quad \left. +\,\pi ^{2}+\frac{187}{72}+O\left( 1-z\right) %
\right] ,  \label{partmass1}
\end{eqnarray}
where 
\begin{equation}
z\equiv 1-\frac{m_{cut}^{2}}{m_{b}^{2}}.
\end{equation}
Dividing eq.\thinspace (\ref{elespec1}) by eq.\thinspace (\ref{partmass1})
under the condition in eq.\thinspace (\ref{vincolo}), i.e. $z=x,$ the
logarithmic terms cancel and the one-loop correction to $C\left( \alpha
_{S}\right) $ in eq.\thinspace (\ref{lei}) results. Note also the
cancellation of the $\pi ^{2}$ terms in taking the ratio of the widths.

Let us now comment on the result represented by eq.\thinspace (\ref{lei}).
The dependence on the non-perturbative effects related to Fermi motion ---
described by the shape function --- cancels in taking the ratio of the
widths. Cancellation occurs also for the CKM matrix element $|V_{ub}|^{2}$
and for the heavy mass power $m_{b}^{5}$, both entering $\Gamma _{0}.$ It is
the cancellation of all these unknown or poorly known quantities that makes
the sum rule quite accurate.

An equation similar to (\ref{lei}), with the replacement $m_{B}\rightarrow
m_{\Lambda _{b}},$ applies also to the hyperon decay 
\begin{equation}
\Lambda _{b}\rightarrow X_{u}+l+\nu .  \label{hyperio}
\end{equation}
The experimental analysis is more difficult in this case, because hyperon
production cross sections are generally much smaller than the corresponding
mesonic ones. The relevance of a combined analysis is that higher-twist
corrections are expected to be different in the two cases (\ref{inizio}) and
(\ref{hyperio}), because for example the $B$-meson has $1/m_{B}$
spin-dependent corrections, which vanish instead in the $\Lambda _{b}$ case 
\cite{memasse}.

In general, we would like to stress the simplicity of the result (\ref{lei}%
). The latter is however non-trivial, as the presence of non-vanishing
perturbative corrections and higher-twist effects indicates. Using only a
general parametrization of the hadronic tensor that describes the decay (\ref
{inizio}), it does not seem possible to derive eq.\thinspace (\ref{lei}).
Let us remark that the prediction (\ref{lei}) does not involve either a
parametrization of the shape function\ or an evaluation of the Mellin
moments of the distributions --- the latter requiring a knowledge of the
spectra in the whole kinematical range. On the experimental side, both the
rates entering eq.\thinspace (\ref{lei}) can be easily measured --- they are
actually measured --- because the background coming from $b\rightarrow c$
transitions is kinematically forbidden\footnote{%
See footnote 1.} \cite{ACCMM,uralt}. \ The sum rule (\ref{lei}) also allows
a consistency check between the electron spectrum computed inside the AC$^{2}
$M$^{2}$ model \cite{ACCMM} \ and the hadron-mass distribution computed
inside the shape-function theory \cite{uralt}. Both these models are
currently used for the experimental determination of $|V_{ub}|.$

To conclude, the experimental confirmation of eq.\thinspace (\ref{lei}) can
provide a check at the $5\%$ level of the theory of the shape function and
of its basic assumptions: infinite-mass limit for the beauty quark,
infinite-energy limit for the light final quark and local parton hadron
duality. Finally, a comparison with accurate experimental data can provide
an estimate of the higher-twist effects.

\bigskip

\begin{center}
{\bf Acknowledgements}
\end{center}

$~~~$

I would like to thank S. Catani and M. Luke for discussions.

\end{document}